\documentclass{aa}

\usepackage{xcolor}
\usepackage{graphicx}
\usepackage{txfonts}
\usepackage{ulem}
\usepackage{amstext}

\begin{document}

   \title{Dark matter capture and annihilation in stars: Impact on the red giant branch tip}

   \author{José Lopes
          \inst{1}
          \and
          Ilídio Lopes\inst{1}
          }

   \institute{Centro de Astrofísica e Gravitação—CENTRA, Departamento de Física, Instituto Superior Técnico—IST,\\
 Universidade de Lisboa—UL, Av. Rovisco Pais 1, 1049-001 Lisboa, Portugal\\ \email{josevlopes@tecnico.ulisboa.pt}
             }

   \date{}

  \abstract
   {While stars have often been used as laboratories to study dark matter (DM), red giant branch (RGB) stars and all the rich phenomenology they encompass have frequently been overlooked by such endeavors.}
   {We study the capture, evaporation, and annihilation of weakly interacting massive particle (WIMP) DM in low-mass RGB stars ($M=0.8$--$1.4~M_{\odot}$).}
   {We used a modified stellar evolution code to study the effects of DM self-annihilation on the structure and evolution of low-mass RGB stars.}
   {We find that the number of DM particles that accumulate inside low-mass RGB stars is not only constant during this phase of evolution, but also mostly independent of the stellar mass and to some extent stellar metallicity. Moreover, we find that the energy injected into the stellar core due to DM annihilation can promote the conditions necessary for helium burning and thus trigger an early end of the RGB phase. The premature end of the RGB, which is most pronounced for DM particles with $m_{\chi}\simeq100~\mathrm{GeV}$, is thus achieved at a lower helium core mass, which results in a lower luminosity at the tip of the red giant branch (TRGB). Although in the current WIMP paradigm, these effects are only relevant if the number of DM particles inside the star is extremely large, we find that for light WIMPs ($m_{\chi}\simeq 4~\mathrm{GeV}$), relevant deviations from the standard TRGB luminosity ($\sim 8\%$) can be achieved with conditions that can be realistic in the inner parsec of the Milky Way.}
   {}


    \titlerunning{Dark matter impact on the red giant branch tip}
    
    \authorrunning{Lopes \& Lopes}
   \maketitle
%
%-------------------------------------------------------------------

\section{Introduction} \label{sec:intro}

The idea of using stars as testing grounds for probing the nature of dark matter (DM) is not new, and many avenues of exploration have been proposed so far \citep[e.g.,][amongst others cited throughout this work]{silk1985,bouquet89,Lopes462,zentner2011,Vincent_2015,lopes2019,lopes2019b}. Developments in stellar modeling and extensive astronomical observations should unlock the full potential of these cosmic laboratories, which, given their unique conditions, represent an important search effort complementary to Earth-based experiments. An excellent example of such laboratories that has often been overlooked are red giant stars. After exhausting the central hydrogen supply, stars will burn hydrogen in a thin shell around an inert helium core that will grow in mass, contract, and warm up in the process. As the core contracts, the hydrogen--rich envelope of the star will expand while simultaneously increasing in luminosity and decreasing in effective temperature. Stars in this phase will thus climb upward on the Hertzsprung-Russel (HR) diagram, forming the well-known red giant branch (RGB), a structure ubiquitous in the color--magnitude diagram (CMD) of galaxies and globular clusters (GC). While the star climbs the RGB, the temperature in the inert core will rise and eventually trigger helium burning, at $T\simeq 10^8~\mathrm{K}$ \citep[e.g.,][]{Kippenhahn:1493272}, which marks the end of the RGB and settles the star in a quiescent phase of core helium-burning at a much lower luminosity. In the case of low-mass stars (i.e., $M \sim 0.5$--$2.0~M_{\odot}$), the core reaches a highly electron--degenerate state before the ignition of helium. In these conditions, the moment at which helium ignites is almost solely dictated by the mass of the growing helium core $M^{\mathrm{He}}_{\mathrm{c}}$, which is approximately independent of the total stellar mass and/or metallicity. The result of this is that all low--mass red giant stars leave the RGB with approximately (depending on the absolute magnitude band) the same luminosity \citep{dacosta1990,lee1993}. The resulting CMD feature is called the tip of the red giant branch (TRGB) and can be identified in GCs, in which stellar populations usually have a well-defined age and metallicity, and in galaxies composed of a complex mixture of stars \citep{barker2004}.

The applications of the TRGB brightness as an absolute CMD reference point in a stellar population are twofold. On one hand, the TRGB brightness can be used as a standard candle to measure distances to local group and distant galaxies ($d \lesssim 10~\mathrm{Mpc}$), serving as an important step in the cosmic distance ladder \citep{sakai1996,maiz2002,igor2006}. On the other hand, if the distance to a given stellar population is well known, its TRGB can be used as testing grounds to constrain new physics \citep{raffelt1992,Catelan_1996}. There are notable examples of this use of the TRGB, such as efforts to constrain the magnetic dipole moment of the neutrino, $\mu_\nu$, and the axion-electron coupling parameter, $g_{ae}$ \citep{Viaux_2013,Capozzi_2020}. Both these parameters are related to nonstandard physical phenomena that increase the rate at which energy is removed from the electron--degenerate core, delaying the onset of helium ignition and thus increasing the absolute brightness of the TRGB.

We are interested in studying the capture and annihilation of weakly interacting massive particle (WIMP) DM in RGB stars and its implications on the absolute brightness of the TRGB. Because they are massive, WIMPs from the galactic halo can be captured by stars, in the center of which they eventually settle by scattering multiple times with baryons as they traverse the stellar plasma \citep[e.g.,][]{Gould1987,PressSpergel1985}. After sinking to the center of the star, captured WIMPs self-annihilate to Standard Model (SM) particles, which, except for neutrinos, promptly deposit their energy within the surrounding plasma. This process acts as an additional energy source whose efficiency is mainly dictated by the WIMP mass $m_{\chi}$ and rate of DM capture. The effects of energy injection from DM annihilation in stars have been studied in many different contexts. \cite{Salati:1987ja} were the first to study the impact of ``cosmion'' DM annihilation in the path of main-sequence (MS) stars, and using polytropic models, argued that these would inflate and move toward the RGB region of the HR. Following studies employing stellar evolution codes \citep{Fairbairn_2008,taoso2008,Iocco_2008,lopes_silk} have not only confirmed these results, but also obtained that DM energy production could extend the lifetimes of stars by counteracting gravitational contraction and contributing to the hydrostatic equilibrium balance. Some authors went even further and found that in ultradense DM halos, characteristic of the early Universe ($z\sim10-50$), the energy released in DM annihilation could completely halt the gravitational collapse in forming protostars, allowing them to maintain a stable configuration devoid of any nuclear reactions \citep{Spolyar_2008, Yoon_2008}. These ``dark stars" would not only profoundly implicate the process of early star formation, but would also be observable at high $z$ because they would be easily distinguishable from nuclear fusion-driven stars. A different (and milder) consequence of these effects was studied by \cite{Casanellas_2011}, who found that GCs embedded in a highly concentrated DM halo could exhibit both an overall younger appearance and a noticeable upward shift in the MS. In a different work, \citet{casa2011} found that a large energy output due to DM annihilation in MS stars could trigger core convection, which would leave an imprint on the asteroseismology of the star. Additionally, the annihilation of captured DM has also been studied in the context of compact stars: the energy released in the degenerate core of white dwarfs and neutron stars can increase the surface temperature of the star, which can be constrained by observations \citep{Moskalenko_2007,Bertone_2008, de_Lavallaz_2010,Kouvaris_2008}. The same reasoning has also been used to constrain a DM signature in the measured heat flow of the Earth \citep{Mack_2007}.

In the approaches that have been devised so far, RGB stars have been almost completely overlooked in the context of energy injection from DM annihilation, which is compelling given the constraining potential of the TRGB. These stars are expected to make such interesting DM laboratories because the release of energy from DM annihilation heats the electron--degenerate core, and this increase in temperature may trigger the early onset of helium ignition, which in turn would decrease the absolute brightness of the TRGB. We use a modified stellar evolution code to study the capture and annihilation of WIMP DM in RGB stars and ultimately, to understand the conditions in which a relevant signature of DM annihilation in the RGB, including the TRGB, can be expected. In the next section we describe the stellar evolution code and the reference RGB stellar models used throughout this work. In section \ref{sec:dm} we study the DM population that accumulates inside RGB stars and the balance of the processes by which it is governed. The formalism of DM energy production is then presented in section \ref{sec:trgb}, which is followed by the results obtained by modeling these effects in RGB stars. A discussion of the results is then presented in section \ref{sec:discussion}, which is followed by the conclusion and some final remarks.

\section{Modeling the tip of the red giant branch}\label{sec:model}

The potential of the TRGB as testing grounds for DM annihilation can only be realized if we are able to accurately predict the star properties, the most relevant of which experimentally is the luminosity $L_{\mathrm{TRGB}}$. The validity of this prediction relies on our ability to robustly model the evolution of the star from the onset of core hydrogen-burning up to the point of helium ignition. While it is true that RGB stars and the physical processes that govern their evolution are relatively simple and well understood, the prediction of the exact moment of helium ignition is still subject to the different subtleties and uncertainties of current stellar modeling. 

Many authors have carried out detailed analyses using different stellar evolution codes to study the theoretical value of the TRGB luminosity, as well as the uncertainties that plague its computation \citep{Valle_2012,Viaux_2013,Serenelli_2017,straniero}. In the most recent of these studies, \citet{straniero} accounted for the uncertainties associated with the most relevant input physics ingredients and through a Monte Carlo propagation obtained a theoretical uncertainty in the computation of the TRGB luminosity of approximately $0.02~\mathrm{dex}$, which is consistent with previous studies. This analysis, however, did not account for important systematic effects such as differences in the adopted numerical algorithms that can increase this uncertainty: for example, \citet{Serenelli_2017} obtained that varying the minimum time step of the model can induce a variation in the luminosity up to $\sim 0.05~\mathrm{dex}$. Moreover, it should be noted that important sources of uncertainty can also arise from bolometric corrections to the TRGB luminosity, which are mostly relevant when theoretical predictions are compared with observations.

Taking advantage of the comprehensive analysis of the uncertainties of $L_{\mathrm{TRGB}}$ carried out by the previous authors, we focused on reproducing the set of RGB reference models by \cite{Serenelli_2017} (hereafter S17), who considered two different stellar evolution codes, and using a set of concordance physical and numerical inputs, obtained a remarkable agreement between them across a wide range of star mass and metallicity. We used the MESA stellar evolution code \citep{Paxton2011, Paxton2013, Paxton2015, Paxton2018, Paxton2019} to model stars in the RGB\footnote{All models used in this work were obtained using MESA release version 12778 -- \url{http://mesa.sourceforge.net/release/2020/03/05/r12778.html}.}, and further modified it to include the phenomenology of DM capture and annihilation (see sec.~\ref{sec:dm}). Because our aim was to study low--mass RGB stars and the generic behavior of the TRGB, we considered broad mass and metallicity ranges similar to S17 and adopted many of their physical assumptions (with some notable exceptions): as in S17, MESA uses the standard radiative opacities from OPAL \citep{Iglesias1993,Iglesias1996}, with low-$T$ data from \citet{Ferguson2005}, and conductive opacities from \citet{Cassisi2007}; nuclear screening is treated according to \citet{Dewitt1973} and \citet{Graboske_1973}, while all thermal neutrino loss-processes are from \citet{Itoh1996}; the nuclear rates are from the JINA Reaclib database \citep{Cyburt2010}, which include H-burning reactions from NACRE \citep{Angulo1999} and updated values for the rate of $^{14}\mathrm{N}(p,\gamma)^{15}\mathrm{O}$ \citep{Imbriani_2005} and the $3\alpha$ reactions \citep{Fynbo_2005} (regarding the latter, while S17 used the older rates from the NACRE collaboration, for the temperature at which the helium flash occurs ($T\sim 10^8~\mathrm{K}$), the differences with the newer rates are negligible); the equation of state (EOS) is a blend of different tables, the relevant tables for our case being OPAL \citep{Rogers2002} and HELM \citep{Timmes2000}. Although these are not the tables used by S17, who used the freeEOS \citep{Cassisi_2003}, they agree well for most relevant cases, including the MS \citep{Vinyoles_2017} and RGB \citep{Viaux_2013}; differently from S17, who used an older metal mixture by \citet[][hereafter GN93]{gn93}, we used the mixture by \citet[][hereafter GS98]{gs98}. For the initial helium abundance, we used the same values as S17, $Y=0.245$ at $Z=0$ and the helium-to-metal enrichment parameter of $\Delta Y/ \Delta Z = 1.4$; convection was modeled according to the standard mixing length theory (MLT) by \citet{cox_1968}, with the solar calibrated parameter $\alpha_{\mathrm{MLT}} = 1.90$, obtained by \citet{capelo_2020} while using the mixture by GS98; just like in \cite{Viaux_2013} and \cite{Serenelli_2017}, our models do not include mass loss during the RGB, nor microscopic diffusion and rotation.

To study the broad range of RGB stars that contribute to the TRGB, we considered stars in the mass range $M =0.8-1.4~M_{\odot}$ and metallicity $Z=0.0001-0.0187$. The helium core mass $M^{\mathrm{He}}_c$ and luminosity at the TRGB $\log L_{\mathrm{TRGB}}$ are shown in Table \ref{table:ref} for some of the reference models. It should also be noted that the time-step controls were chosen such that near the TRGB, the time step was approximately $\Delta t \sim 1~\mathrm{kyr}$.

\begin{table}[t]
\small
\centering
\caption{Properties of selected reference models (i.e., with no DM energy production) at the TRGB. Age is relative to the ZAMS. For a detailed explanation of the physics assumptions, see text.}
\begin{tabular}{lllrll}
 $Z$ & $Y$ & $M/M_{\odot}$ & Age (Gyr)& $M^{\mathrm{He}}_c$ & $\log \left(L_{\mathrm{TRGB}}/L_{\odot}\right)$ \\ \hline \hline
0.0001 & 0.245 & 0.8 & 12.48 & 0.494 & 3.230 \\
 &  & 1.0 & 5.76 & 0.486 & 3.193 \\
 &  & 1.4 & 1.91 & 0.455 & 3.015 \\ \hline
0.0040 & 0.251 & 0.8 & 17.23 & 0.477 & 3.363 \\
 &  & 1.0 & 7.67 & 0.474 & 3.351 \\
 &  & 1.4 & 2.39 & 0.471 & 3.335 \\ \hline
0.0187 & 0.271 & 0.8 & 28.51 & 0.469 & 3.402 \\
 &  & 1.0 & 12.57& 0.467 & 3.401  \\
 &  & 1.4 & 3.61 & 0.466 & 3.394 \\ \hline
\end{tabular}
\label{table:ref}
\end{table}

As expected, the similarity in the values of $\log L_{\mathrm{TRGB}}$ for different $M$ and same $Z$-$Y$ pair show that the luminosity at the tip of the RGB is very weakly dependent on the mass of the star. The same cannot be inferred in regard to the metallicity because for higher $Z$, $L_{\mathrm{TRGB}}$ can be more than twice higher than for lower metallicity, which can hinder the determination of the TRGB for a composite stellar population. Overall, the TRGB luminosity and helium core mass obtained across the mass and metallicity ranges agree with the results in S17 within the quoted margins of uncertainty. We also compared our models with results from other authors \citep{Valle_2012,Viaux_2013,straniero}, and they are consistent within what is expected from models with different numeric and physics inputs. To conclude this section, we stress that while we tried to be as thorough as possible in respect to the choice of appropriate input physics, the purpose of this exercise was to establish a baseline of canonical RGB models (i.e., with no DM annihilation) that are consistent with the current picture of RGB modeling. Having obtained these reference models, we can safely study the effects of DM annihilation, which at first order are expected to be independent of the physics assumptions adopted in the baseline models.

\section{Number of DM particles in low--mass RGB stars}\label{sec:dm}

In the standard picture of self-annihilating DM, the equation that governs the evolution of the number of WIMPs inside a star is given by
\begin{equation}
\frac{\mathrm{d}N_{\chi}}{\mathrm{d}t} = C - EN_{\chi} - AN_{\chi}^2,
\label{eq:budget}
\end{equation}
which describes the balance of the three different processes that contribute to total budget of $N_{\chi}$ (the full expressions for these quantities can be found in the appendix). Capture, $C$, occurs when a DM particle from the galactic halo interacts within the stellar plasma and scatters to a velocity lower than the local escape speed thus being unable to break away from the gravitational pull of the star. It is defined by the properties of the galactic halo such as the local DM density $\rho_{\chi}$, the mass and scattering cross-section of the DM particle , $m_{\chi}$  and $\sigma_{\chi}$ respectively, and the structure and composition of the star. Evaporation, $E$, is the process where a particle that belongs to the DM population of the star scatters with matter from the stellar hot plasma and acquires enough energy to pull away from the gravitational potential. It is therefore highly dependent on the conditions inside the star. Finally, when two captured DM particles interact, they annihilate to SM particles that promptly loose their energy within the surrounding plasma. The annihilation rate, $\Gamma_{\mathrm{A}} \equiv \frac{1}{2}AN_{\chi}^{2}$, is defined by the thermally averaged annihilation cross section $\langle \sigma v \rangle$ and the radial distribution of the DM population in the star $n_{\chi}(r)$.

The magnitude of the effects caused by energy production in DM annihilation is directly related to the number of particles that populate the center of the star. Our aim in this section is thus to understand how the processes of capture, evaporation, and annihilation behave during the RGB, and what their role is in the rate of DM energy production up to the TRGB.

When a star evolves through the different stages that lead up to the TRGB, the different coefficients in Eq.~\ref{eq:budget} vary as the structure of the star changes, and so does the total budget of particles that can potentially contribute to $\Gamma_{\mathrm{A}}$. While at first sight this could mean that it would be necessary to track the evolution of $N_{\chi}$ from the early stages of stellar evolution and solve Eq.~\ref{eq:budget} accordingly, in most practical cases, the number of particles is almost always at an equilibrium, that is, $\mathrm{d}N_{\chi}/\mathrm{d}t = 0$. When this is the case, $N_{\chi}(t)$ is completely defined by the properties of the star at each instant $t$ and is independent of its past history. This equilibrium can be characterized by two distinct regimes in which the balance in $N_{\chi}$ is maintained between capture and either evaporation or annihilation. For lower $m_{\chi}$, most DM-nucleon interactions will give the WIMP the energy necessary to escape the star, meaning that evaporation is more efficient than annihilation in counterbalancing capture. However, $E$ decays exponentially with increasing $m_{\chi}$, which quickly renders evaporation negligible when compared to annihilation, which overcomes the former in maintaining balance with capture. While there is no clear-cut boundary between the two regimes, it is useful to define an evaporation mass, $m_{\chi}^{\mathrm{evap}}$, above which $E$ is negligible, and below which $N_{\chi}$ drops dramatically because evaporation renders any DM effects negligible. Although this quantity has often been computed for the Sun \citep[e.g.,][]{Busoni_2013}, there is no obvious reason why the result should be valid in stars as different as the red giant stars considered here. For this reason, and following the recipe in \citet{Busoni_2013}, we computed $m_{\chi}^{\mathrm{evap}}$ for a benchmark model with $M=1.0M_{\odot}$ and $Z=0.0187$ at four different points in evolution: at a stage during the MS (representative of the present Sun); when the star begins to ascend the RGB, just before the star undergoes the RGB bump, and when the star reaches the TRGB. Capture, annihilation, and evaporation were computed according to the expressions shown in the appendix, assuming a spin-independent cross section $\sigma_{\chi,\mathrm{SI}} = 10^{-39}~\mathrm{cm}^2$, thermally averaged cross section $\langle \sigma v \rangle_{\mathrm{ann}} = 3 \times 10^{-26}~\mathrm{cm}^3\mathrm{s}^{-1}$, and the solar values for the stellar velocity $v_0 = 220~\mathrm{km s}^{-1}$ and DM halo velocity distribution dispersion $v_\mathrm{d} = 270~\mathrm{km s}^{-1}$. These parameters are also used throughout the rest of this work, unless stated otherwise.

\begin{figure}[!t]
    \centering
    \includegraphics[width=\columnwidth]{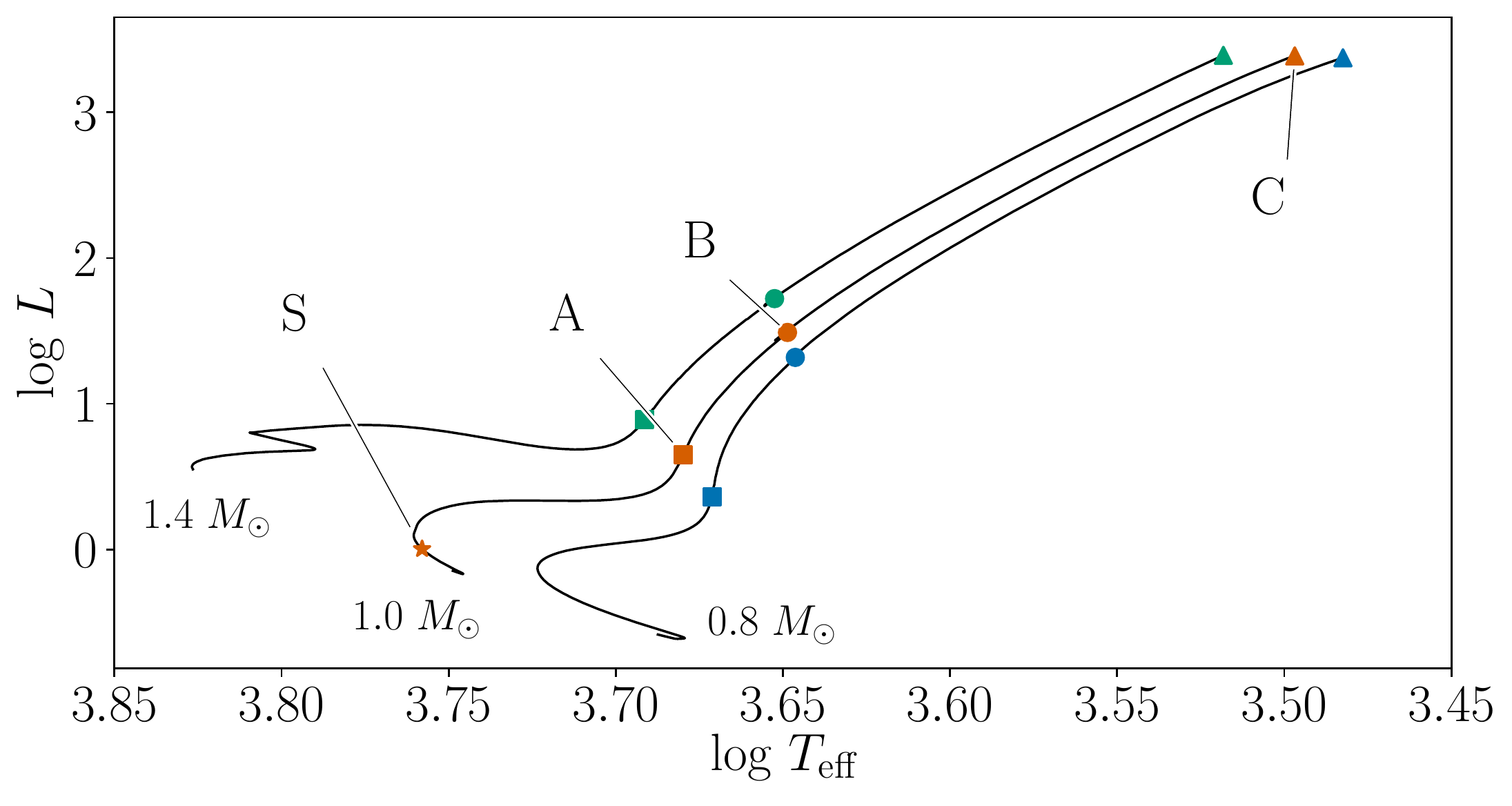} \hfill 
   \caption{HR diagram with the evolution of $M=0.8,1.0$ and $1.4~M_{\odot}$ stars starting from the ZAMS up to the TRGB. Relevant benchmark models (S, A, B, and C) are marked in the evolution tracks (see text for more details).}
    \label{fig:hr_evap}
\end{figure}

\begin{table}[]
\caption{Evaporation mass $m_{\chi}^{\mathrm{evap}}$ for a $M=1.0~M_{\odot}$ star during different phases of evolution: S in the MS, A at the start of the RGB, B just before the RGB bump, and C at the TRGB (see Fig.~\ref{fig:hr_evap}).}
\begin{tabular}{lll|lll}
 & & $m_{\chi}^{\mathrm{evap}}~[\mathrm{GeV}]$ & & & $m_{\chi}^{\mathrm{evap}}~[\mathrm{GeV}]$ \\ \hline \hline
S & $\bigstar$ & 3.0 & B & $\bullet$ & 1.9 \\
A & $\blacksquare$ & 2.3 & C & $\blacktriangle$ & 1.3\\
\end{tabular}
\label{tab:mevap}
\end{table}

The results in Table \ref{tab:mevap} (see also Fig.~\ref{fig:hr_evap}) show that when the star is in the MS, the evaporation mass is $m_{\chi}^{\mathrm{evap}} = 3.0~\mathrm{GeV}$, which is in agreement with similar computations carried out for the Sun \citep[e.g.,][]{Busoni_2013,Garani_2017}. More importantly, our results show that the evaporation mass slightly decreases as the star evolves through the RGB. This decrease is related to the core contraction that occurs during the RGB, which has two separate consequences: on one hand, it is harder for the DM particles to escape the deepened gravitational potential; on the other hand, the core contraction promotes further clustering of the DM population, which enhances  self-annihilation.  We carried out the same analysis for stars with $M=0.8M_{\odot}$ and $M=1.4M_{\odot}$ and found that the results in Table \ref{tab:mevap} do not change much ( $\sim10\%$ at most). This result is especially interesting because it shows that stars in the uppermost region of the RGB can probe DM with lower masses (e.g., $m_{\chi} \sim 2~\mathrm{GeV}$) that are not accessible with the Sun. Moreover, it should be noted that these results were obtained assuming the solar neighborhood local DM density $\rho_{\chi} = 0.4~\mathrm{GeV cm}^{-3}$ \citep[e.g.,][]{Catena2010}, which means that if we instead considered regions in which the density is higher, such as in the center of the galaxy, $m_{\chi}^{\mathrm{evap}}$ should be lower than what we obtained here.

The results obtained for $m_{\chi}^{\mathrm{evap}}$ hence show that in the stellar mass range of interest, we can safely neglect evaporation during the RGB for particles with $m_{\chi} \gtrsim 3~\mathrm{GeV}$. In this regime, if capture and annihilation are maintained in equilibrium, the number of DM particles is simply given by 

\begin{equation}
    N_{\chi,\mathrm{eq}} = \sqrt{\frac{C}{A}},
    \label{eq:neq}
\end{equation}

which can be independently evaluated at each instant $t$.

Using equation \ref{eq:budget} while neglecting evaporation and without assuming equilibrium\textup{} a priori, we computed $N_{\chi}(t)$ for a number of models with fixed metallicity and different mass. We considered a broad spectrum of DM mass values $m_{\chi} = 5,100,\text{and }1000~\mathrm{GeV}$ that cover the plausible WIMP mass range and satisfy the conditions obtained in Table \ref{tab:mevap}. 

\begin{figure}[!t]
    \centering
    \includegraphics[width=\columnwidth]{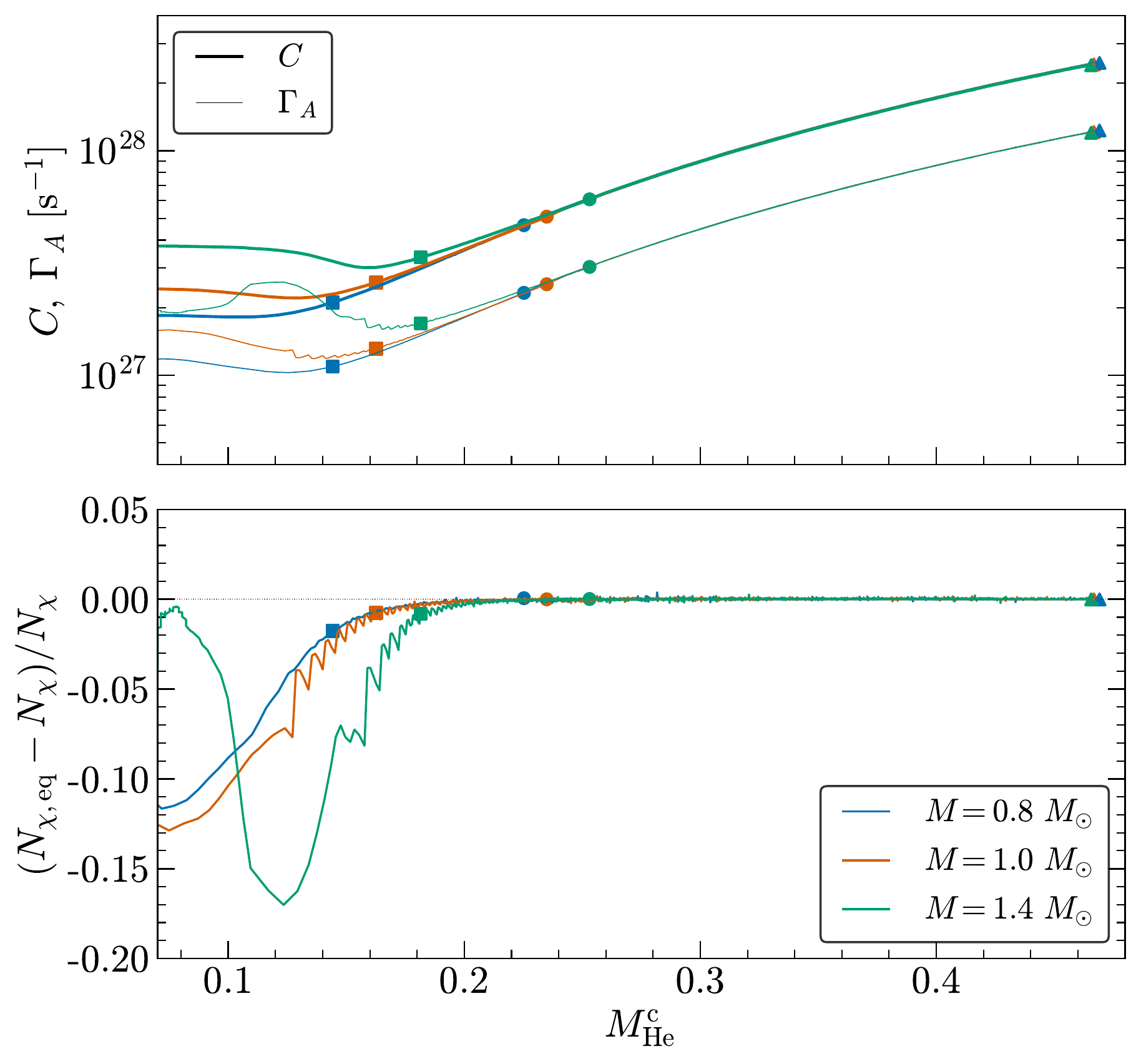} \hfill 
   \caption{Top: $C$ and $\Gamma_{A}$ for $m_{\chi} = 5~\mathrm{GeV}$ in stars with $M=0.8,1.0$ and $1.4~M_{\odot}$ during the RGB. Bottom: Relative difference between $N_{\chi}$ and the number of WIMPs when capture and annihilation are in equilibrium (see Eq.~\ref{eq:neq}). The legend for the markers is the same as in Fig.~\ref{fig:hr_evap}.}
    \label{fig:cap_ann}
\end{figure}

\begin{figure}[!t]
    \centering
    \includegraphics[width=\columnwidth]{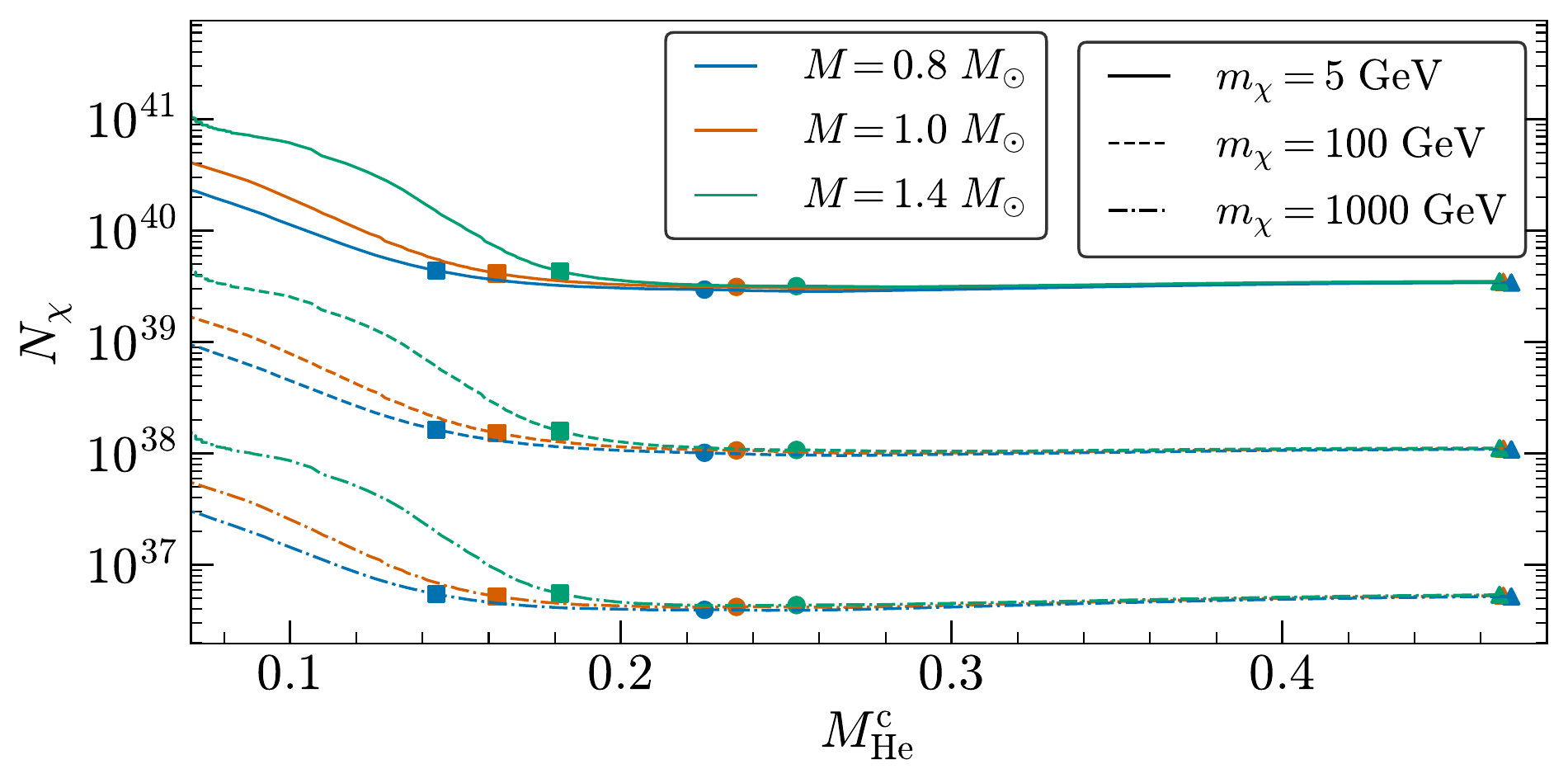} \hfill 
   \caption{Total number of DM particles inside different stars with fixed metallicity ($Z=0.0187$) during the RGB. The legend for the markers is the same as used in Fig.~\ref{fig:hr_evap}.}
    \label{fig:wimp_n}
\end{figure}

The results are shown in Figs.~\ref{fig:cap_ann} and \ref{fig:wimp_n} as a function of the size of the helium core $M_{\mathrm{He}}^{\mathrm{c}}$, which is representative of the stellar evolution during the RGB and allows a direct comparison between stars of different mass. The results in Fig.~\ref{fig:cap_ann} show that capture and annihilation are in equilibrium during most of the RGB, that is, $C\approx 2\Gamma_A$. This is true not only for $m_{\chi} = 5~\mathrm{GeV}$, which is the case shown in Fig.~\ref{fig:cap_ann}, but also for $m_{\chi}$ as high as $1000~\mathrm{GeV}$. There is a moment prior to the RGB where equilibrium is not observed, however, corresponding to the initial phase of core contraction after the end of the MS, which enhances DM annihilation in the core, and momentarily tips the scale in favor of the annihilation rate.  

This effect is also observed in Fig.~\ref{fig:wimp_n}, where as a consequence of the increase in annihilation, the number of DM particles decreases as the star approaches the RGB, before reaching the equilibrium given by Eq.~\ref{eq:neq}. As stated before, this behavior is observed for all the considered values of $m_{\chi}$. Moreover, the number of particles during the RGB is approximately independent of the stellar mass. The reason is that during the RGB and especially in its upper part, most DM capture ($\sim 95\%$), and virtually all annihilation occurs within the helium core, which, as stated before, is independent of the total mass of the star. This means that similarly to $L_{\mathrm{TRGB}}$, the number of DM particles during the RGB is independent of $M$ (in the considered mass range).

\begin{figure}[!t]
    \centering
    \includegraphics[width=\columnwidth]{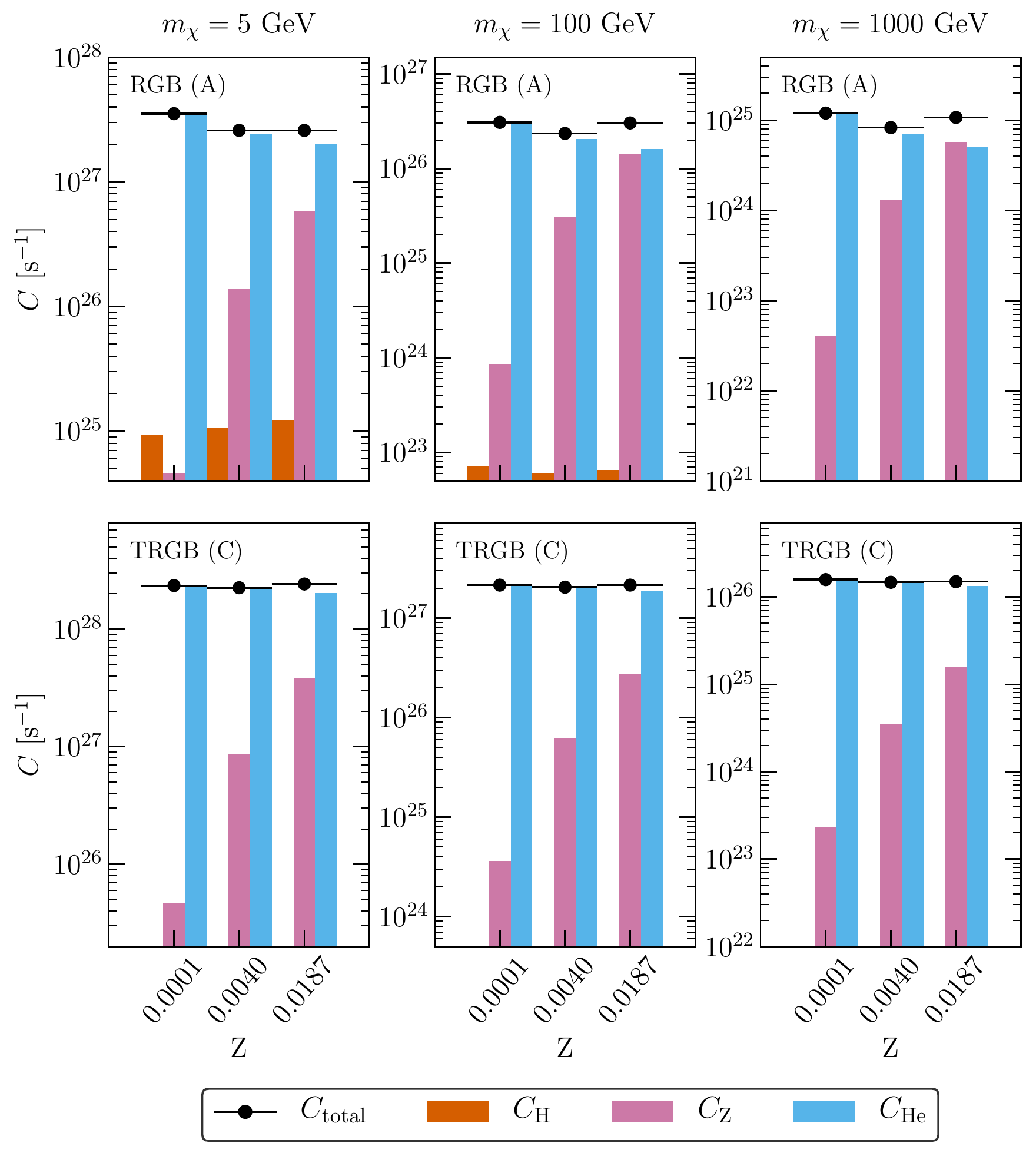} \hfill 
   \caption{Capture rate for different $m_{\chi}$ in a $1.0~M_{\odot}$ star with different metallicity values. We show the total capture and the capture due to H, He, and all the metals, i.e., elements of higher mass than He.}
    \label{fig:cap_metal}
\end{figure}

\begin{figure}[!t]
    \centering
    \includegraphics[width=\columnwidth]{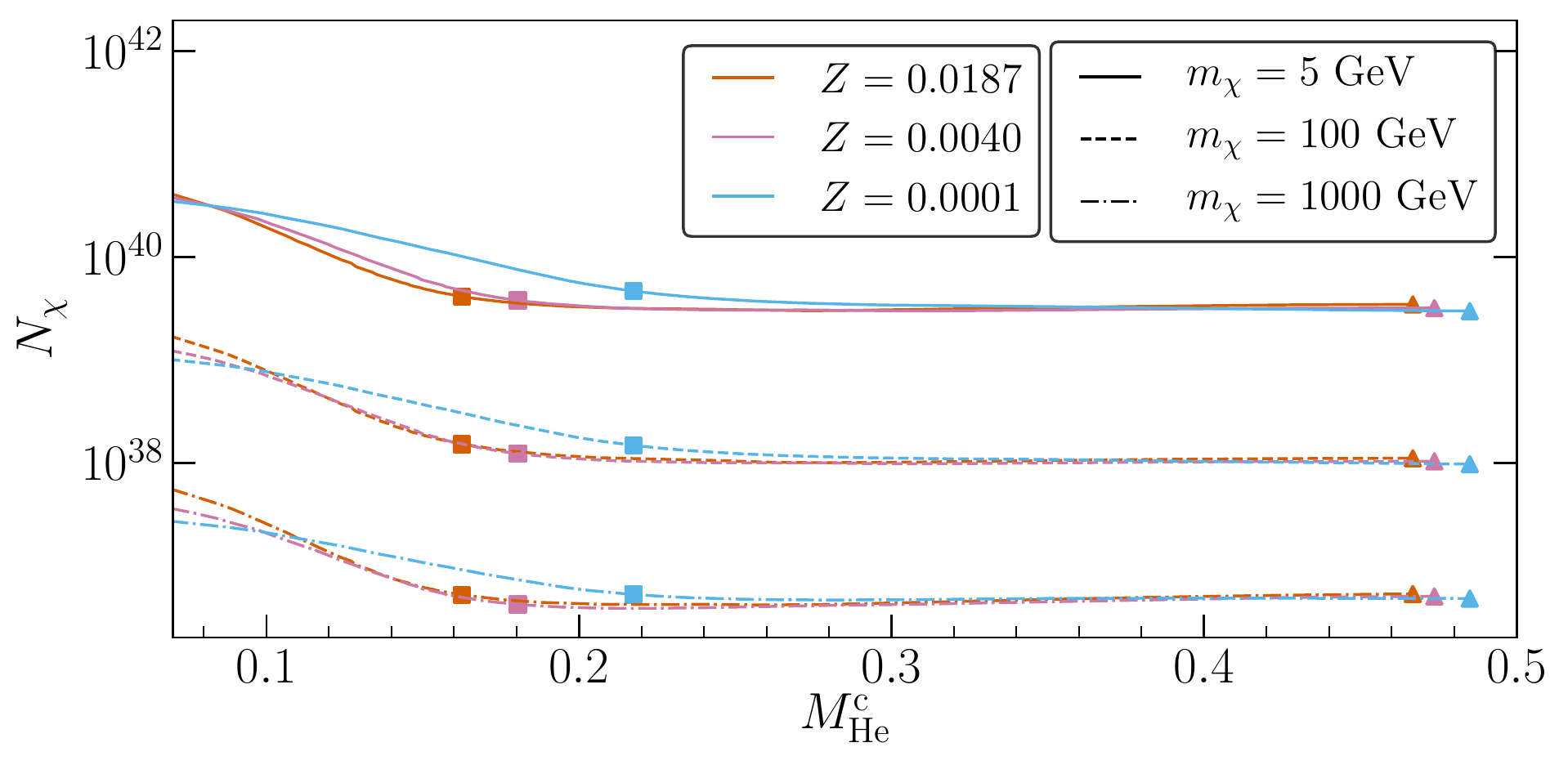} \hfill 
   \caption{Same as in Fig.~\ref{fig:wimp_n}, but for stars with the same mass ($M=1.0~M_{\odot}$) and different metallicity.}
    \label{fig:wimp_n_metal}
\end{figure}

To evaluate the effect of the different metallicities on $N_{\chi}$ , we compared the capture rate due to hydrogen ($C_{\mathrm{H}}$), helium ($C_{\mathrm{He}}$), and higher-mass elements ($C_{\mathrm{Z}}$) at the beginning of the RGB and at the TRGB (Fig.~\ref{fig:cap_metal}). The results show that hydrogen capture is negligible when compared to the other contributions, and that in most cases, DM capture is essentially driven by scatterings with helium (which is expected given that most capture occurs in the He core). The exception to this rule occurs for higher $m_{\chi}$ at the beginning of the RGB, where there is an important $C_{\mathrm{Z}}$ contribution, which can be explained by the fact that higher-mass DM particles are more prone to be captured by higher-mass chemical elements. More importantly, the results in Fig.~\ref{fig:cap_metal} show that while there is a noticeable difference in $C$ for models with different $Z$ at the beginning of the RGB, by the time the star reaches the TRGB, these differences are mitigated by the growing helium core, which is independent of $M$ and is where virtually all capture takes place. This is also true for the total number of particles as a function of $Z$ shown in Fig.~\ref{fig:wimp_n_metal}, which along with Fig.~\ref{fig:wimp_n} shows that $N_{\chi}$ is not only mostly independent of $M$ and $Z$, but also approximately constant as the star climbs the RGB. This result is not limited to the parameters assumed here ($\sigma_{\chi}$ and $\rho_{\chi}$) as it holds as long as capture and annihilation are in equilibrium, which is the case for $\left(\sigma_{\chi}\rho_{\chi}\right) \gtrsim 10^{-45}~\mathrm{GeV~cm}^{-1}$.

It should be noted that the results shown in this section were obtained using models in which the energy balance does not account for the additional term due to DM annihilation. Nonetheless, while it is true that this contribution can have an important role in the TRGB (as shown in the following sections), its effect on the structure of the star and on $N_{\chi}(t)$ during the RGB is negligible.

\section{Energy production and luminosity from DM annihilation}\label{sec:trgb}

After entering the grasp of the gravitational pull of a star, a captured DM particle is bound to undergo subsequent interactions before settling in some type of thermodynamical equilibrium within the stellar plasma. For RGB stars with $M\approx 1~M_{\odot}$, taking into account that most capture occurs within the star helium core, this process of thermalization occurs on a timescale not longer than \citep{Iocco_2008}
\begin{equation}
    \tau_{\mathrm{th}} \approx 1.1~\mathrm{yr}\times\left(\frac{m_{\chi}}{1000~\mathrm{GeV}} \frac{10^{-37}~\mathrm{cm}^2}{\sigma_{\chi,\mathrm{He}}}\right),
\end{equation}
where $\sigma_{\chi,\mathrm{He}}$ is the DM-scattering cross section with helium (see Eq.~\ref{eq:dm_cs}). For the parameters considered in this work ($m_{\chi} = 5-1000~\mathrm{GeV}$ and $\sigma_{\chi,\mathrm{SI}} = 10^{-39}~\mathrm{cm}^2$), $\tau_{\mathrm{th}}$ is always much smaller than the minimum time step of the model, $\Delta t \sim 1~\mathrm{kyr}$, which means that for our purposes, we can assume that DM particles thermalize instantly after capture. Regarding the thermodynamical equilibrium, the distribution on which the DM settles after thermalization is defined by the interactions of the DM particle and its surroundings. In the case of RGB stars, DM particles cluster very strongly within the inert helium core, the temperature and density profiles of which are nearly constant. For example, for a $m_{\chi} = 5~\mathrm{GeV}$ particle in an RGB star with $1~M_{\odot}$, 90\% of the DM population is contained within the first $0.01~M_{\odot}$ of the star, which accounts for a maximum variation of 2\% and 9\% in the temperature and density, respectively. Higher-mass DM particles will even cluster strongly in the center of the star, which reinforces our argument. After thermalization, we can therefore assume that DM particles occupy a spherical region with uniform temperature and density, in which case the DM distribution is well described by a Maxwell-Boltzmann distribution given by
\begin{equation}
n_{\chi}(r) = N_0 \exp \left(-\frac{r^2}{r_{\chi}^2}\right), \quad r_{\chi} \equiv \left( \frac{3kT_{\mathrm{c}}}{2\pi G\rho_{\mathrm{c}} m_{\chi}} \right)^{\frac{1}{2}}
\label{eq:n_chi}
\end{equation}

where $N_0$ is a normalization constant defined by the total number of particles, $T_\mathrm{c}$ and $\rho_\mathrm{c}$ are the central temperature and density, respectively, and the length scale $r_{\chi}$ can be interpreted as the radius of the DM distribution. The energy produced by DM annihilation at a given radius $r$ per unit mass and per unit time is thus given by

\begin{equation}
\varepsilon_{\chi}(r) =f_{\nu} \frac{\langle \sigma v \rangle n_{\chi}^2(r) m_{\chi}}{\rho(r)},
\label{eq:en_prod}
\end{equation} 

where $\rho(r)$ is the local nuclear matter density, and $f_{\nu}$ is a dimensionless factor that accounts for the energy lost through annihilation byproducts (such as neutrinos) that escape the star before loosing any energy within the stellar plasma. Differently from most assumptions made so far, the factor $f_{\nu}$ is highly model dependent. We assumed that the amount of energy carried away from the degenerate helium core by neutrinos is very small and thus $f_\nu \approx 1$. The total luminosity due to DM annihilation is given by the integral of Eq.~\ref{eq:en_prod},
\begin{equation}
L_{\chi} = 4 \pi \int_0^{R_*} \rho(r) \varepsilon_{\chi}(r) r^2\mathrm{d}r=f_{\nu} m_{\chi} C,
\label{eq:lum_chi}
\end{equation} 
where the last equivalence, which can be obtained by using Eq.~\ref{app:annihilation} and assuming capture--annihilation equilibrium (see Eq.~\ref{eq:neq}), shows that both $L_{\chi}$ and $\varepsilon_{\chi}$ are independent of the annihilation cross section $\langle \sigma v \rangle$. 

It should be noted that DM particles that have thermalized and settled in the stellar interior can scatter with the hot plasma and act as an additional heat transfer mechanism, which can have an effect on the structure of the star \citep[e.g.,][]{gilliland86}. While this additional energy transport mechanism is mostly irrelevant in the core of an RGB star (where the electron degeneracy renders conduction the most efficient heat transfer mechanism), it can have a significant effect during the preceding hydrogen core burning phase, when, for example, a small change in the heat transfer balance can dictate the onset of convection \citep[e.g.,][]{casanellas15,lopes2019,raen21}. The efficiency of the DM heat transfer is directly related to the number of DM particles captured inside the star, that is, $N_{\chi}$, and thus is considerably suppressed for DM with a non-negligible annihilation cross section, which is the central ingredient of this work. For this reason we did not take DM energy transport into account and focused on the effects caused by DM annihilation only.

\section{Effects of DM annihilation on the RGB}\label{sec:dm_and_trgb_results}

\begin{figure}[!t]
    \centering
    \includegraphics[width=\columnwidth]{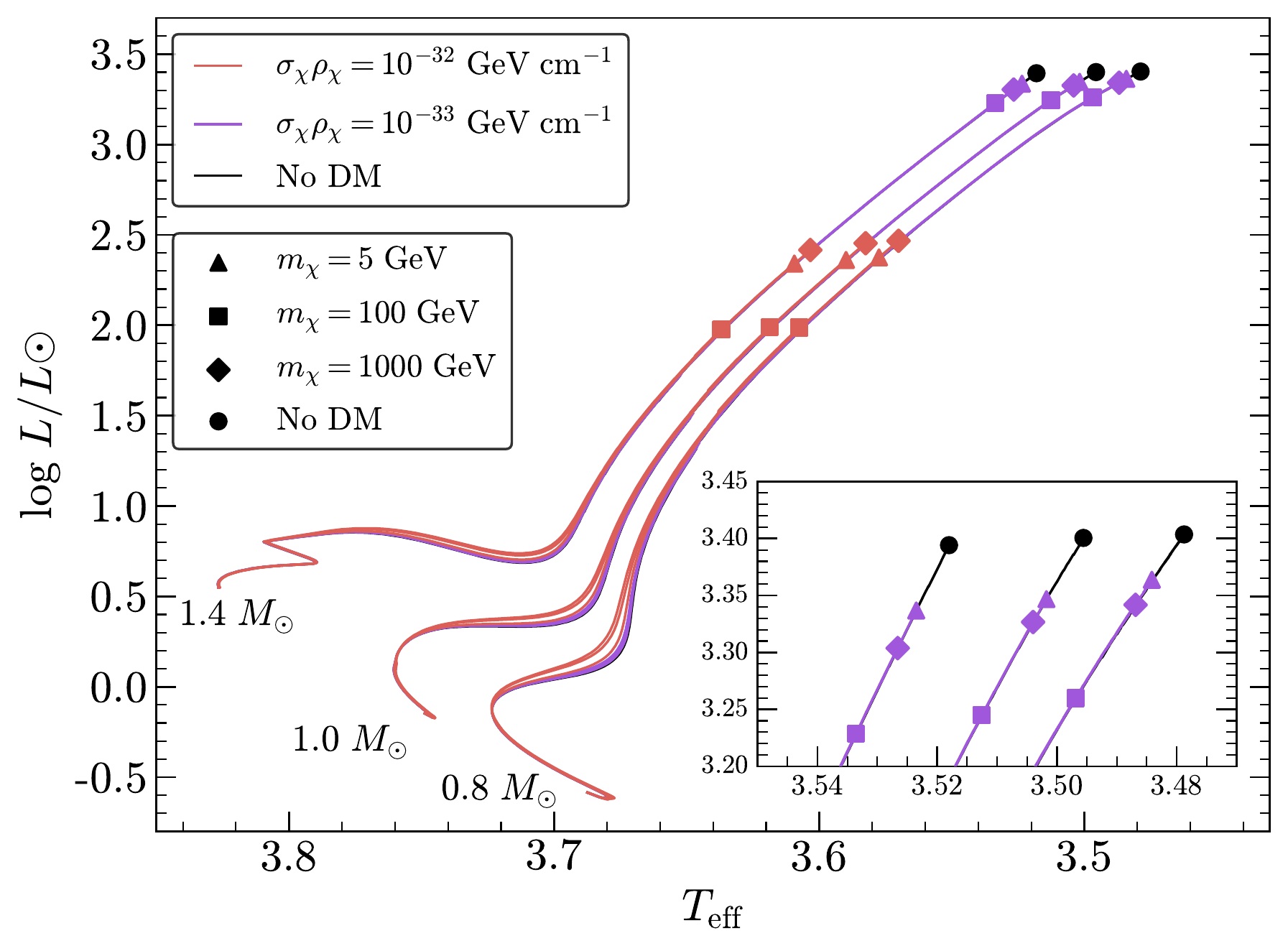} \hfill 
   \caption{HR diagram for low--mass RGB stars ($Z=0.0187$) with DM energy production for different DM models ($m_{\chi}$ and $\sigma_{\chi}\rho_{\chi}$). The respective benchmark models (i.e., with no DM) are also shown. All models begin in the ZAMS and evolve until the ignition of helium.}
    \label{fig:hr_dm}
\end{figure}

To study the effects of DM in the evolution of low--mass RGB stars, we modified the MESA stellar evolution code\footnote{The subroutines and files necessary to reproduce the results in this work can be found in doi:\url{10.5281/zenodo.4769667}.} taking into account DM energy production as described by Eq.~\ref{eq:en_prod}. The number of DM particles at each time step was obtained by computing capture and annihilation and solving Eq.~\ref{eq:n_chi}. The DM energy term as well as the corresponding partial derivatives with respect to the structural variables of the star was then computed and included in the set of differential equations that govern stellar evolution. 

Figure~\ref{fig:hr_dm} shows the HR diagram for a set of RGB stars ($M=0.8-1.4~M_{\odot}$, $Z=0.0187$) embedded in DM halos with different properties ($\sigma_{\chi}\rho_{\chi}=10^{-33}-10^{-32}~\mathrm{GeV~cm}^{-1}$), starting from the zero-age MS (ZAMS) and stopping when helium ignition occurs, that is, at the TRGB. Because the capture rate is proportional to $\sigma_{\chi}\rho_{\chi}$, we use this parameter to present our results. Non-DM related physics inputs are the same as in the benchmark models described in Sec.~\ref{sec:model}.

As expected, the main effect caused by DM energy production in RGB stars is the early onset of helium ignition and end of the RGB phase, which results in an overall lower luminosity at the TRGB. The reason for this is that a localized net energy source in the center of the highly degenerate helium core increases the central temperature of the star, which  promotes the conditions necessary to trigger the helium runaway reaction. The higher the DM energy output $\varepsilon_\chi$ , or equivalently, the higher the DM luminosity $L_{\chi}$ , the earlier the conditions for helium ignition are met. It should be noted that the cause--effect relation of the DM energy injection and  temperature increase is intrinsically linked to the degenerate nature of the inert helium core. This is in contrast with nondegenerate stars such as the Sun, in which the energy released in nuclear reactions acts as a ``thermostat'' by counterbalancing contraction. 

\begin{figure}[!t]
    \centering
    \includegraphics[width=\columnwidth]{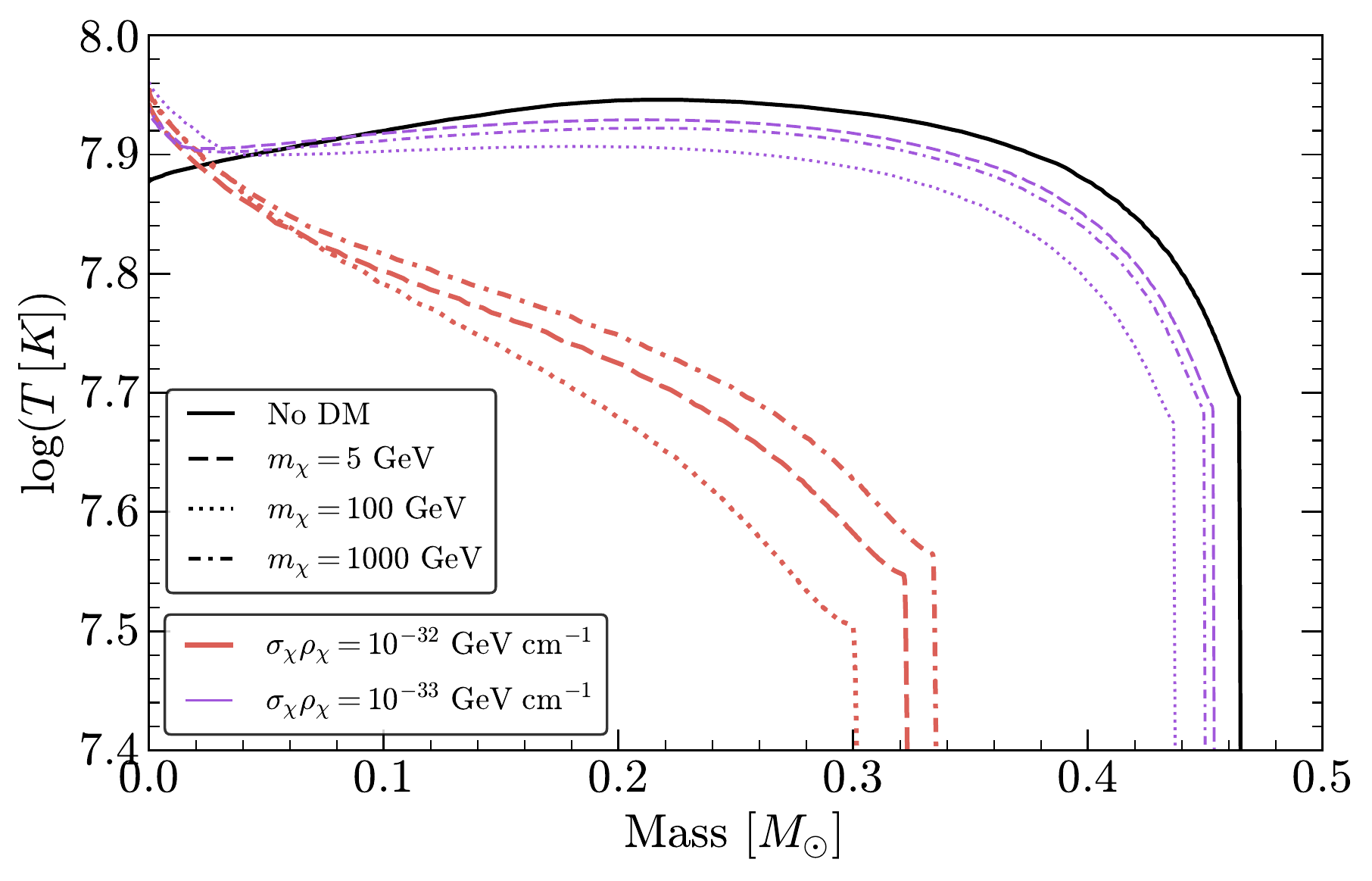} \hfill 
   \caption{Temperature profile for a solar mass RGB star ($Z=0.187$) just before the onset of helium burning for different DM models ($m_{\chi}$ and $\sigma_{\chi}\rho_{\chi}$).}
    \label{fig:t_profile}
\end{figure}

Another effect of DM energy production in RGB stars that is shown in Fig.~\ref{fig:t_profile} concerns the region of the core in which helium ignites first: while in stars without DM, helium ignition occurs off-center due to energy loss by plasmon neutrinos \citep[e.g.,][]{Kippenhahn:1493272}, the energy released by DM annihilation (maximum at $r=0$ by definition) ensures that the conditions necessary for helium nuclear fusion are first met in the center of the core. Moreover, while for $\sigma_{\chi}\rho_{\chi}=10^{-33}~\mathrm{GeV~cm}^{-1}$ the temperature profile exhibits a second maximum coincident with the region of ignition in the standard case ($M\sim0.2~M_{\odot}$), for $\sigma_{\chi}\rho_{\chi}=10^{-32}~\mathrm{GeV~cm}^{-1}$ , the triggering of helium burning at a much lower core mass prevents the outer regions of the core from reaching temperatures at which neutrino loss is significant, which results in a monotonically decreasing temperature profile. It should be noted that this result has no meaningful effect on our results or conclusions because the runaway reaction in any case spreads rapidly within the degenerate helium core \citep{Kippenhahn:1493272}.

\begin{figure}[!t]
    \centering
    \includegraphics[width=\columnwidth]{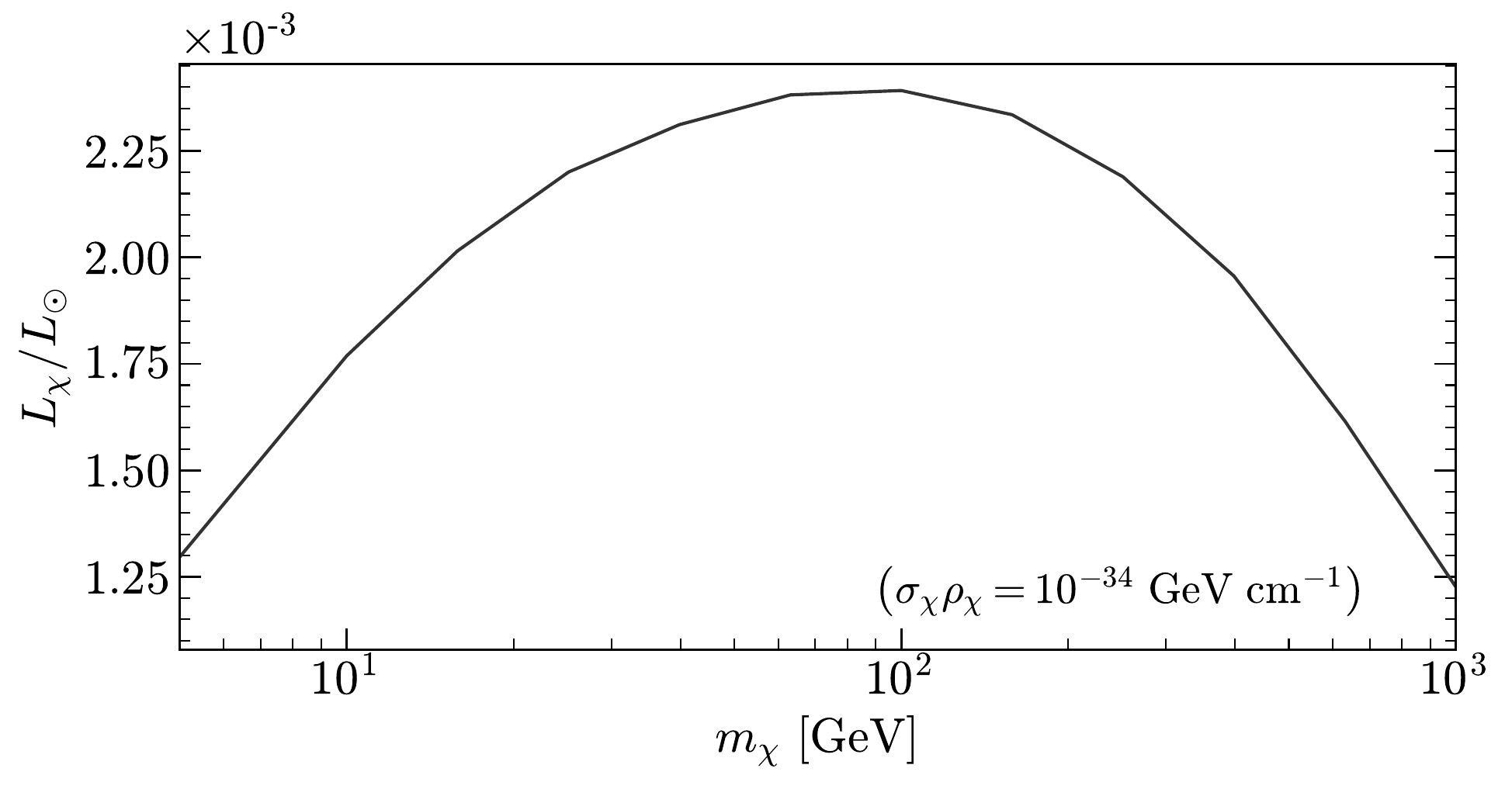} \hfill 
   \caption{DM luminosity as a function of the DM mass at the reference value of $\log~L/L_{\odot}=2.0$ (see Fig.~\ref{fig:l_chi_evol} for reference).}
    \label{fig:l_chi_mass}
\end{figure}

Additionally, the results in Fig.~\ref{fig:hr_dm} show that while $L_{\mathrm{TRGB}}$ is lower for higher values of $\sigma_\chi \rho_\chi$, which follows directly from Eq.~\ref{eq:lum_chi}, the relationship between $m_{\chi}$ and the TRGB luminosity warrants a more detailed analysis. First, the relation between the TRGB luminosity and the DM mass is not monotonic. This can be understood from the fact that the DM luminosity $L_{\chi}$ increases with the mass of the DM particle and with the number of annihilation events per unit time. These in turn depend on the DM capture rate (see Eq.~\ref{eq:lum_chi}), which decreases with $m_{\chi}$, meaning that there is a mass value for which $L_{\chi}$ is highest. This behavior is shown in Fig.~\ref{fig:l_chi_mass}, which shows $L_{\chi}$ as a function of $m_{\chi}$ for RGB stars approximately in the same position of the HR diagram ($\log~L/L_{\odot}=2.0$). $L_{\chi}$ clearly has a maximum for $m_{\chi}\simeq100~\mathrm{GeV}$, and thus we expect the largest departures in $L_{\mathrm{RGB}}$ from the standard non-DM case in this DM mass region.

Moreover, when we compare the models with $m_{\chi}=5~\mathrm{GeV}$ and $1000~\mathrm{GeV,}$  the answer to which of these cases corresponds a lower luminosity at the TRGB is not independent of $\rho_\chi \sigma_\chi$: for $\sigma_\chi \rho_\chi = 10^{-33}~\mathrm{GeV~cm}^{-1}$, models with $m_{\chi} = 5~\mathrm{GeV}$ have a higher TRGB luminosity than models with $m_{\chi} = 1000~\mathrm{GeV}$ (triangles are higher than diamonds). On the other hand, for $\sigma_\chi \rho_\chi = 10^{-32}~\mathrm{GeV~cm}^{-1}$, the situation is the reverse (triangles are lower than diamonds). 

\begin{figure}[!t]
    \centering
    \includegraphics[width=\columnwidth]{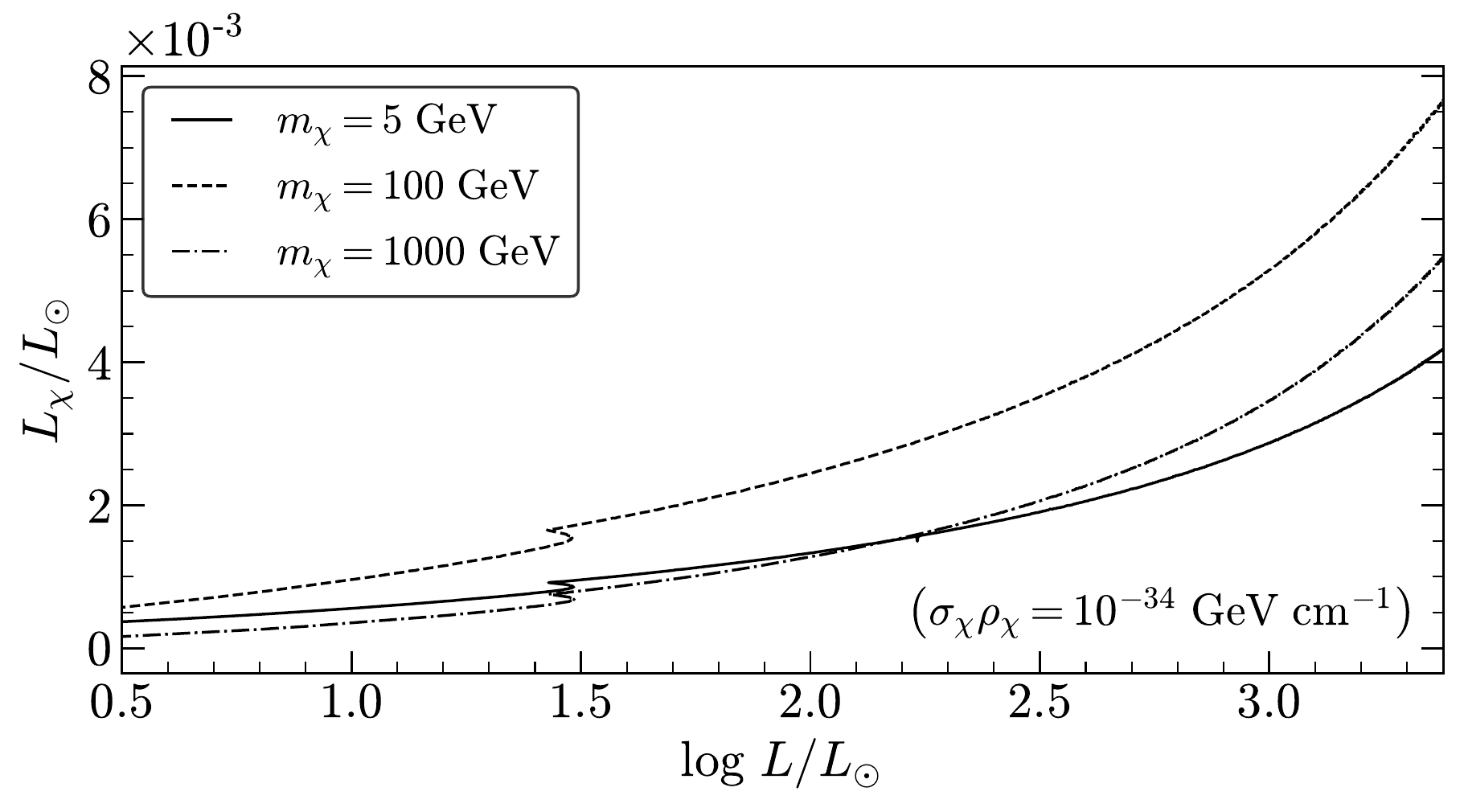} \hfill 
   \caption{DM luminosity as a function of the total luminosity, which here represents the evolution of the star during the RGB (see Fig.~\ref{fig:hr_dm} for reference), for different values of DM mass. The wiggle present in all cases at $L_{\chi}/L_{\odot}\simeq1.5$ corresponds to the RGB bump.}
    \label{fig:l_chi_evol}
\end{figure}

This behavior can be understood by studying Fig.~\ref{fig:l_chi_evol}, where we show how $L_{\chi}$, which in our regime is equivalent to $f_\nu m_\chi C$, increases during the RGB for different values of $m_{\chi}$ (unless stated otherwise, all results shown hereafter correspond to an RGB star with $M=1.0~M_{\odot}$ and $Z=0.0187$). As the star climbs the RGB and the helium core grows, not only does the region that is mainly responsible for capture increase in size, but DM particles from the galactic halo that were otherwise inaccessible due to dynamical constraints (such as large velocities) also become susceptible to the deepening gravitational potential of the star. These are the main factors contributing to the increase in capture that is observed during the RGB, as we showed in Fig.~\ref{fig:cap_ann}. However, the rate at which $C$ increases is different for each value of $m_{\chi}$. The results in Fig.~\ref{fig:l_chi_evol} show that at the start of the RGB, $L_{\chi}$ is higher for $m_{\chi} = 5~\mathrm{GeV}$ than for $m_{\chi} = 1000~\mathrm{GeV}$. However, as the helium core increases and is able to capture both faster and higher-mass DM particles, capture for $m_{\chi} = 1000~\mathrm{GeV}$ has a steeper increase, and there is a moment at which the resulting luminosity overcomes the case with $m=5~\mathrm{GeV}$. It should be noted that while the results in Figs.~\ref{fig:l_chi_mass}--\ref{fig:l_chi_evol} correspond to specific values of $\sigma_{\chi}\rho_{\chi}$, the point they illustrate, that is, the relationship between $L_{\chi}$ and $m_{\chi}$, can be generalized to other values this parameter.

\begin{figure}[!t]
    \centering
    \includegraphics[width=\columnwidth]{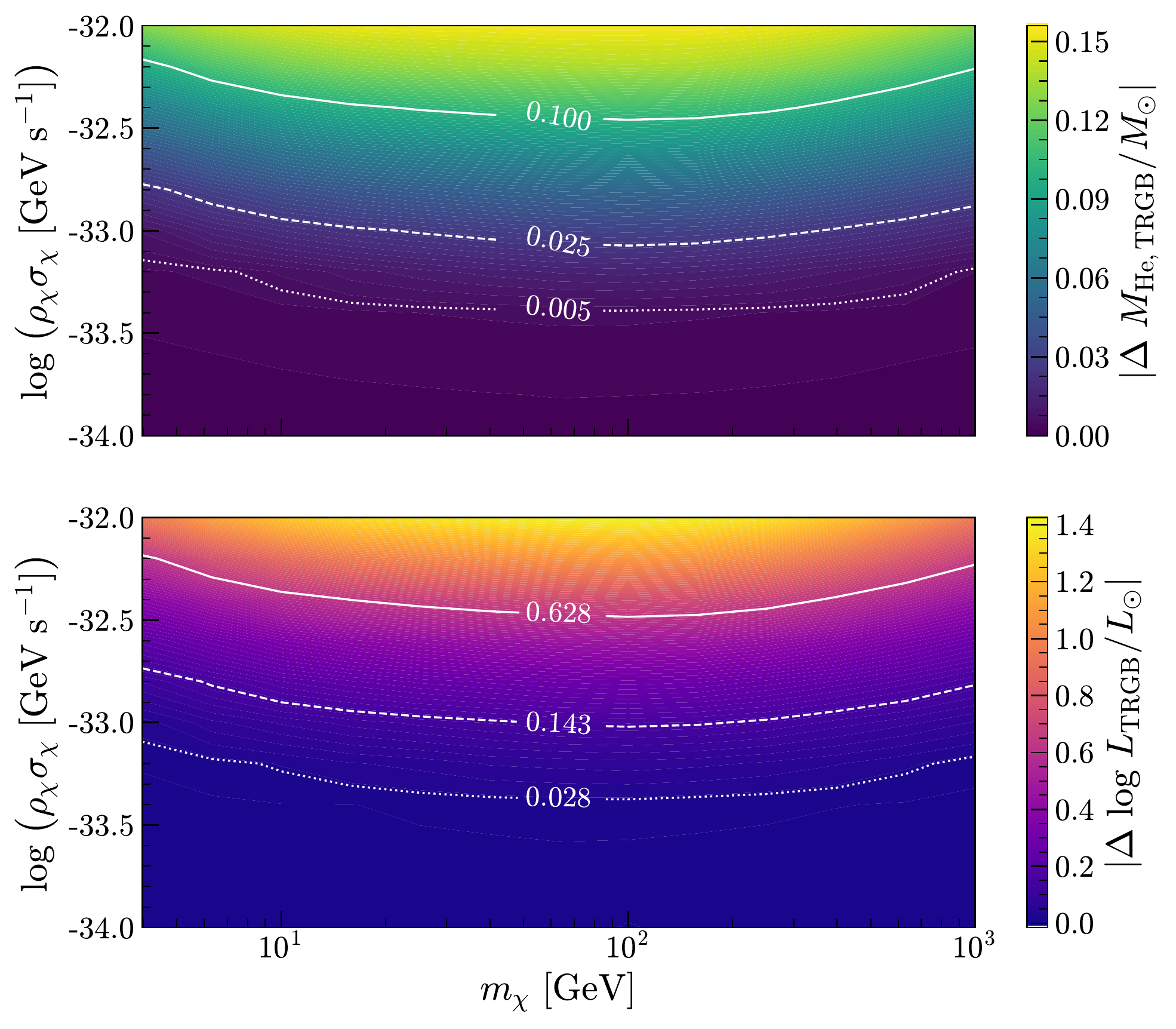} \hfill 
   \caption{Absolute variation of the TRGB helium core mass (top panel) and luminosity (bottom panel) to the standard non-DM model as a function of $m_{\chi}$ and $\sigma_{\chi}\rho_{\chi}$. The values are shown in modulus for readability, but they are always negative in both panels (see text for more details). Contours with the same line style are related according to Eq.~\ref{eq:m_l_rel}.}
    \label{fig:trgb_color}
\end{figure}

Finally, in Fig.~\ref{fig:trgb_color} we collect the results obtained for the mass of the helium core at the TRGB and corresponding luminosity as a function of $m_{\chi}$ and $\sigma_{\chi}\rho_{\chi}$. More specifically, we show the absolute difference of $M^\mathrm{He}_{\mathrm{c}}$ and $\log L_{\mathrm{TRGB}}$ to the standard case without DM. It should be noted that while the differences are shown as absolute values, they are in fact always negative. The results shown in Fig.~\ref{fig:trgb_color} somewhat summarize what has been discussed so far in this section: the effects of DM annihilation in $L_{\mathrm{TRGB}}$ (and $M^\mathrm{He}_{\mathrm{c}}$) increase with $\sigma_{\chi}\rho_{\chi}$, and there is a mass region in which these effects are highest, centered around $m_{\chi}\simeq 100~\mathrm{GeV}$ . This region, as described before, is determined by the balance of the number of annihilation events and the energy released in each event. Furthermore, we also show the contours for relevant variations in helium core mass at the TRGB.  While the contours in the top panel are in fact the models for which the displayed variation is observed, the corresponding contours in the lower panel were derived from the latter, according to the semi-empirical relation \citep{Serenelli_2017}
\begin{equation}
    \frac{\partial \log L_{\mathrm{TRGB}}/{L_{\odot}}}{\partial \log M_{\mathrm{c}}^{\mathrm{He}}/M_{\odot}} \sim 6.
    \label{eq:m_l_rel}
\end{equation}
The result that each $\Delta M_{\mathrm{c}}^{\mathrm{He}}$--$\Delta \log L_{\mathrm{TRGB}}$ pair of contours (curves with the same line style) approximately outlines the same region in the $\rho_\chi \sigma_\chi$--$m_{\chi}$ parameter space confirms the validity of the relation in Eq.~\ref{eq:m_l_rel}. Moreover, the fact that a relation that was derived from standard stellar evolution considerations also holds for effects caused by DM confirms the robustness of the results. In terms of relative variations: models with $\Delta M_{\mathrm{c}}^{\mathrm{He}} = 0.005, 0.025$ and $0.100~M_{\odot}$ (1.1\%, 5.4\%, and 21.4\% lower core mass, respectively) will have a  corresponding luminosity $\delta L_{\mathrm{TRGB}} \approx 6.4\%, 27.9\%,$ and $76.4\%$ dimmer, respectively (note that the last case represents a luminosity $ \text{about four}$ times lower than the standard case). In particular, the contour for $\Delta M_{\mathrm{c}}^{\mathrm{He}}/M_{\odot} = 0.005$ shown in Fig.~\ref{fig:trgb_color} is especially relevant because estimations by \citet{Catelan_1996} that took theoretical and observational considerations into account show that the helium core mass at the TRGB should not exhibit a deficit larger than this value.

\section{Discussion}\label{sec:discussion}

The results obtained in Sec.~\ref{sec:dm_and_trgb_results} were presented in terms of the parameter $\sigma_{\chi}\rho_{\chi}$ instead of the individual properties that define it, that is, the DM--nucleon cross section and DM density. One of the main reasons that led us to do so is that if the criteria for capture--annihilation equilibrium and ``instant" thermalization are satisfied, the results can readily be generalized for different values of $\sigma_{\chi}$ and $\rho_{\chi}$. However, it is important to consider each of the quantities that compose the parameter $\sigma_{\chi}\rho_{\chi}$ individually and understand where and when (or even if) it is realistic to expect such DM properties or conditions. For example, in order to observe the effect of a $100~\mathrm{GeV}$ particle with $\sigma_{\chi}\rho_{\chi} \simeq 10^{-33}~\mathrm{GeV cm}^{-1}$ in an RGB star in our vicinity, $\rho_{\chi} = 0.4~\mathrm{GeV cm}^{-3}$ \citep[e.g.,][]{Catena2010}, the DM particle would need to have $\sigma_{\mathrm{SI}}\simeq 3 \times 10^{-33}~\mathrm{cm}^2$, which is many orders of magnitude above the limits imposed by current direct detection experiments, $\sigma_{\mathrm{SI}}\lesssim 10^{-46}~\mathrm{cm}^2$ \citep[][XENON1T]{Aprile_2018}. On the other hand, assuming the same value of $\sigma_{\chi}\rho_{\chi}$ while respecting the latter limit implies that the star would need to be embedded in a DM density of $\rho_{\chi} \sim 10^{13}~\mathrm{GeV cm}^{-3}$. Although many questions still remain regarding this quantity, we know that the most DM--rich environments in the universe are expected to be found within the central regions of large galaxies such as the Milky Way. Results from numerical simulations extrapolated to the center of the galaxy yield a DM density value that in the innermost 1 (0.01) pc of the MW can range from $\rho_{\chi} \sim 10^{5}~(10^{8})~\mathrm{GeV cm}^{-3}$ to $\rho_{\chi} \sim 10^{6}~(10^{10})~\mathrm{GeV cm}^{-3}$, depending on how the adiabatic contraction \citep{blumenthal_1986,Gnedin_2004} and density spike due to the central black hole \citep{Gondolo_1999} are considered. This means that even in the most extreme scenario, we would not be able to see any relevant signature in a $1~M_{\odot}$ RGB star due to a $100~\mathrm{GeV}$ DM particle that does not contradict direct detection experiments in the usual WIMP paradigm. While this is the general case for particles higher in mass than $10~\mathrm{GeV}$, direct detection experiments are not as stringent in the low--end mass spectrum: for $m_{\chi} \simeq 4~\mathrm{GeV}$, the lowest-mass DM particle for which evaporation is negligible, direct searches place the limits at $\sigma_{\mathrm{SI}}\lesssim 10^{-42}~\mathrm{GeV}$ \citep{Aprile_2019,Agnes_2018}. In these conditions, an RGB star in a region with $\rho_{\chi} \sim 10^{9}~\mathrm{GeV cm}^{-3}$, which is not unrealistic at the galactic subparsec scale, will have a TRGB helium core $\sim2\%$ lighter and a corresponding luminosity $\sim8\%$ lower than in the standard no-DM case. 

The potential of constraining annihilating DM with the TRGB looks even more promising when we take into account that the DM velocity distribution used in this work, an MB distribution with $v_d =270~\mathrm{km s}^{-1}$ boosted to a frame with $v_\odot=220~\mathrm{km s}^{-1}$, often results in overly conservative estimates of the capture rate and, hence, of the DM effects on the star. As we approach the galactic center, it is not only expected that the DM velocity distribution will deviate from the MB approximation, skewing toward lower velocities \citep{Lopes_2021}, but also that the stellar velocity in the galactic frame is lower than the value observed for the Sun, $v_\odot=220~\mathrm{km s}^{-1}$. These two effects will boost the DM capture rate. As shown in \citet{Lopes_2021}, a simple but important improvement on the treatment of the DM halo distribution can be achieved by assuming an MB with the most probable speed given by the Jeans equation. Using this approximation for a typical 1.0 $M_{\odot}$ RGB star (model A in Fig.~\ref{fig:hr_evap}) located at 5 pc away from the center of the MW, with a circular velocity defined by the mass enclosed by the stellar orbit ($v_* \simeq 66~\mathrm{km s}^{-1}$), we obtained that the capture rate is enhanced by more than a factor of 2. The results for the effects on the TRGB shown here should therefore be regarded as conservative estimates, especially for cases with higher densities, which are associated with the central regions of the galaxy.

Regardless of the possibility of encountering DM conditions that fit the $\sigma_{\chi}\rho_{\chi}$ parameter region studied here, the expectations of observing a clear-cut signature of DM annihilation on the TRGB are intrinsically limited by the associated phenomenology. On the one hand, the fact that the change in luminosity is a function of the DM density, which in turn changes throughout the galaxy, means that $L_{\mathrm{TRGB}}$, as a property of a stellar population rather than a single star, would not exhibit a well-defined gap, but instead a gradient that would range from the standard non-DM value (for stars in lower DM density regions) to a lower luminosity from stars that are located closer to the galactic center. This radial gradient in $L_{\mathrm{TRGB}}$ could itself be an indication of a DM density gradient at play, but it would be hard to distinguish from other non-DM effects. Similarly, because DM annihilation will trigger the early end of the RGB at a lower luminosity, it is spectroscopically impossible within an RGB population to distinguish stars that are on the brink of a DM-induced helium flash from stars with no DM that will continue to climb the RGB. This is in contrast with phenomena that in contrast to DM annihilation increase the luminosity of the TRGB, such as those studied in \citet{Viaux_2013}, and create signatures such as particularly luminous low--mass RGB stars that effectively increase the TRGB of the target stellar population. Taking these limitations into account, and given a stellar population with a well-defined RGB, one way to search for the manifestation of the effects studied here would be to search for a deficit in high-luminosity low--mass RGB stars. 

The premature end of the RGB caused by DM annihilation can also have important consequences for the subsequent phase of evolution of the star, the horizontal branch. Some of these stars, the so called RR Lyrae, exhibit pulsations that follow a well-defined period--luminosity relation, and for this reason, they are often used as standard candles to measure extragalactic distances \citep[e.g.,][and references therein]{rrl_rev}. These relations depend on the properties of the star such as the total helium abundance or the size of the helium core, both of which are defined by the moment at which the helium flash occurs, and thus are susceptible to the effects of DM energy injection. It as been shown that a discrepancy between the computed pulsations and observations might be caused by a systematic underestimation of the HB helium core mass \citep{catelan1992}. Interestingly, DM annihilation would deepen this discrepancy by inducing an even smaller $M_{\mathrm{c}}^{\mathrm{He}}$ , thus rendering these stars as potential probes of the effects studied here. This subject warrants a separate and dedicated analysis.

\section{Conclusion}\label{sec:conclusion}

We studied the capture, annihilation, and consequent energy production of WIMP-like DM in low--mass RGB stars. We also studied evaporation, the process in which a DM particle trapped inside the star escapes by scattering to a velocity higher than the local escape speed, and found that for stars at the beginning of the RGB, it is negligible for DM particles with $m_{\chi}\gtrsim 2.3~\mathrm{GeV}$. This limit, which defines the lowest-mass DM particle that can be probed in a given star, decreases as the star climbs the RGB, reaching the minimum value of $m^{\mathrm{evap}}_{\chi}\simeq 1.3~\mathrm{GeV}$ at the TRGB. This is an important result because it shows that RGB stars can capture (and thus probe) DM particles of lower mass than those that can be probed in the Sun, for which $m^{\mathrm{evap}}_{\chi}\simeq 3.0~\mathrm{GeV}$. Focusing on particles of higher mass than these limits ($m_{\chi}\geq4~\mathrm{GeV}$) and RGB stars with $M=0.8$--$1.4~M_{\odot}$, we found that both DM capture and annihilation rates increase during the RGB, while maintaining an equilibrium that renders the total number of DM particles inside the star roughly constant during this phase. When we assume capture--annihilation equilibrium, which is the case in most relevant scenarios ($\rho_{\chi} \sigma_{\chi} \gtrsim 10^{-45}~\mathrm{GeV~cm}^{-1}$), the total number of particles during the RGB is solely defined by the capture and annihilation, meaning that it is independent of the previous evolution phases of the star. Moreover, we find that the number DM particles during the RGB is mostly independent of stellar mass and metallicity, which is explained by the fact that most capture ($\sim 86\%$--$99\%$) occurs within, and due to the helium core.

In the second part of this work, we studied the effects of DM self--annihilation on the structure and evolution of RGB stars. Using a modified stellar evolution code, we find that the energy produced in DM annihilation can precipitate the end of the RGB by heating the degenerate core and promoting the necessary conditions ($T\sim10^8~\mathrm{K}$) for the onset of helium burning. We also find that this DM--triggered helium burning first occurs in the center of the star, whereas in the standard picture of stellar evolution, it occurs off-center due to neutrino energy losses. The resulting premature helium flash occurs at a lower helium core mass, which in turn results in a lower overall luminosity at the TRGB. While the intensity of this effect is proportional to $\rho_{\chi}$ and $\sigma_{\chi}$, its relation with $m_{\chi}$ is dictated by a balance of the capture (decreases with $m_{\chi}$) and the energy released in each annihilation event (increases with $m_{\chi}$). Taking this into account, we find that the effect is strongest for particles with $m_{\chi}\simeq 100~\mathrm{GeV}$. From studying this phenomenology within a wide range of DM properties, we find that in the case of a DM particle with $m_{\chi}=4~\mathrm{GeV}$ and $\sigma_{\mathrm{SI}}\simeq 10^{-42}~\mathrm{cm}^2$ (which is allowed by current direct detection limits), an RGB star in a region with $\rho_{\chi} \sim 10^{9}~\mathrm{GeV cm}^{-3}$ (which is not unrealistic in the inner parsec of the MW) will have a helium core at the TRGB that is $\sim2\%$ smaller and a corresponding TRGB luminosity that is $\sim8\%$ lower. These effects are not limited to WIMPs with these properties, and can be generalized to any DM model with a similar corresponding annihilation rate (in this case, for $m_{\chi}=4~\mathrm{GeV}$, $\Gamma_A \simeq 10^{34}~\mathrm{s}^{-1}$), including models that are not as constrained as the standard WIMP. Moreover, an effect of this magnitude can be relevant not only for the properties of the TRGB of a given stellar population, but also for the premature HB pulsating stars (RR Lyrae) that result from the DM-induced core helium flash.

\begin{acknowledgements}
      We would like to thank Joseph Silk for the discussions that motivated part of this work and Thomas Lacroix for very insightful comments and contributions to the galactic halo phase space distribution discussion. We would also like to thank the anonymous referee for all the comments and suggestions that helped improving this work. J.L. acknowledges financial support from Funda\c c\~ao para a Ci\^encia e Tecnologia (FCT) grant No. PD/BD/128235/2016 in the framework of the Doctoral Programme IDPASC---Portugal. I.L. and J.L. thank the Funda\c c\~ao para a Ci\^encia e Tecnologia (FCT), Portugal, for the financial support to the Center for Astrophysics and Gravitation (CENTRA/IST/ULisboa) through Grant Project~No.~UIDB/00099/2020 and, in the case of I.L., also Grant No. PTDC/FIS-AST/28920/2017. 
\end{acknowledgements}

\bibliographystyle{aa}
\bibliography{AA_2021_40750_biblio}

\begin{appendix}
\section{Formalism}\label{app:formalism}

In this section we outline the full expressions we used to compute the capture, evaporation, and annihilation rates. These processes have been thoroughly studied in the literature. Here we closely follow the formalism presented in \citet{Gould1987},\citet{gould1987b},\citet{Garani_2017}, and \citet{Busoni_2017}. 

The DM capture is the process in which a DM particle with mass $m_{\chi}$ from the galactic halo, with velocity $u$ at infinity, interacts with a nucleon of the stellar plasma, with mass $m_i$ at a shell with radius $r$, and acquires a final velocity lower than the local escape speed, $v_{\mathrm{esc}}(r)$. It is given by a sum over the $i$ elements that contribute to the capture process,

\begin{equation}
C = \sum_i \int_0^{R_*} \mathrm{d}r 4\pi r^2 \int_0^{\infty} \mathrm{d}u \frac{f(u)}{u} w \Omega_{i}(0,v_{\mathrm{esc}};w),   
\end{equation}

where $R_*$ is the radius of the star, $f(u)$ is the DM halo speed distribution in the stellar frame normalized to the local DM number density $\rho_{\chi}/m_{\chi}$, $w$ is the velocity of the DM particle before scattering,
\begin{equation}
    w(r) = \sqrt{u^2 + v_{\mathrm{esc}}^2(r)},
\end{equation}

and $\Omega_{i}(0,v_{\mathrm{esc}};w)$ is the rate at which a DM particle with velocity $w$ interacting with the target nucleus will scatter to a velocity between 0 and $v_{\mathrm{esc}}(r)$.

Evaporation is the process inverse to capture, where a captured DM particle scatters to a velocity higher than the local escape speed, and following the same logic, is given by

\begin{equation}
E = \sum_i \int_0^{R_*} \mathrm{d}r 4\pi r^2 \int_0^{v_{\mathrm{esc}}(r)} \mathrm{d}w f_{\chi}(w) \Omega_{i}(v_{\mathrm{esc}},\infty;w)    
,\end{equation}

where $f_{\chi}(w,r)$ is the DM phase space distribution in the star normalized to the number density $n_{\chi}(r)$. It should be noted that in the definitions of capture and evaporation, we assume that the stellar plasma is optically thin, which is valid for DM particles with a small cross section. If this were not the case, both definitions would include a radial suppression factor \citep[e.g.,][]{Busoni_2017}. The rate $\Omega_{i}(v_{\mathrm{min}},v_{\mathrm{max}};w)$ present in both capture and evaporation definitions is given by

\begin{equation}
    \Omega_{i}(v_{\mathrm{min}},v_{\mathrm{max}};w) = \int_{v_{\mathrm{min}}}^{v_{\mathrm{max}}}\!\mathrm{d}v\, R_i(w\rightarrow v) \left| F_i(\Delta E)\right|^2\,, \\
\end{equation}

where $\left| F_i(\Delta E)\right|^2$ is the nuclear form factor for nucleus $i$ that exchanges an amount of energy $\Delta E \simeq m_{\chi}(w^2-v^2)/2$ with the DM particle and can be approximated by
\begin{equation}
    \left| F_i(\Delta E)\right|^2 = \exp \left( - \frac{\Delta E}{E_i}\right)\,, \quad E_i = \frac{3 \hbar^2}{2 m_i R_i^2}.
\end{equation}

The rate of scattering from a velocity $w$ to $v$, assuming a velocity-independent and isotropic cross section $\sigma_{\chi,i}$ , is given by
\begin{equation}
\begin{split}
    R_i(w\rightarrow v) = &\frac{16}{\sqrt{\pi}} \frac{\mu_{i,+}^4}{u_i^3(r)}\frac{v}{w}n_i(r) \sigma_{\chi,i} \\
    &\times \int_0^{\infty} \mathrm{d}t \int_0^{\infty} \mathrm{d}s~t~e^{-u^2/u_i^2(r)} H(s,t,w,v)
\end{split}
    \label{app:r_rate}
\end{equation}

where 
\begin{equation}
    \mu_{i,+} \equiv \frac{\mu_i+1}{2},\quad \mu_i = \frac{m_\chi}{m_i},
\end{equation}
the most probable speed of the target nucleus is defined by the stellar temperature
\begin{equation}
    u_i(r) = \sqrt{\frac{2 T(r)}{m_i}},
\end{equation}
$n_i(r)$ is the target nucleus number density and the cross section with nucleus $i$ (assuming only spin-independent interactions) is given by 
\begin{align}
\sigma_{\chi,i} = A_i^2 \beta_i^2 \sigma_{\chi,\mathrm{SI}}~\left( \sigma_{\chi,\mathrm{SD}} = 0 \right), \qquad \beta_i = \left(\frac{m_{\chi}+m_\mathrm{p}}{m_\mathrm{p} m_{\chi}}\right)\left(\frac{m_i m_{\chi}}{m_{\chi}+m_i}\right),
\label{eq:dm_cs}
\end{align}
where $A_i$ is the nucleus mass number and $m_{\mathrm{p}}$ is the mass of the proton.

The variables $s$ and $t$ represent the velocity center of mass and the velocity of the DM particle before scattering in the center of mass frame, respectively, and the integrals in Eq.~\ref{app:r_rate} are carried over,
\begin{equation}
    H(s,t,w,v) \equiv \Theta(w-|s-t|)\Theta(s+t-w)\Theta(v-|s-t|)\Theta(s+t-v).
\end{equation}
In the case of capture, given that the speed of the incoming DM particle is typically much higher than the plasma thermal velocity, $w \gg u_i(r)$, we can take the limit $T(r)\rightarrow 0,$ which significantly simplifies Eq.~\ref{app:r_rate}. On the other hand, the evaporation process is intrinsically connected to the velocity of the target nucleus, and thus the plasma temperature has to be accounted for, rendering the computation of evaporation a complex task. 

Finally, the DM annihilation rate is defined as
\begin{equation}
\Gamma_{\mathrm{A}} \equiv \frac{1}{2}N_{\chi}^2 A,
\end{equation}
where $A$ is the self-annihilation coefficient, given by 
\begin{equation}
    A = \frac{\int_{0}^{R_*} \langle \sigma v \rangle n_{\chi}^2(r) r^2\mathrm{d}r}{\left( \int_{0}^{R_*} n_{\chi}(r) r^2 \mathrm{d}r \right)^2},
    \label{app:annihilation}
\end{equation}
where $\langle \sigma v \rangle$ is the thermally averaged annihilation cross section.
\end{appendix}

\end{document}